%% file: EuclidTP_1yr.tex
\documentclass[a4paper,11pt]{spie}  %>>> use this instead for A4 paper
\usepackage{amsmath,amsfonts,amssymb}
\usepackage{graphicx}
\usepackage[colorlinks=true, allcolors=blue]{hyperref}

\newcommand{\taue}{\ensuremath{\tau_{\rm e}}}
\newcommand{\tph}{\ensuremath{t_{\rm ph}}}
\newcommand{\eP}{\ensuremath{P}}
\newcommand{\um}{\textmu{}m}
\newcommand{\us}{\textmu{}s}
\newcommand{\Euclid}{\textit{Euclid}}
\newcommand{\Ltwo}{\ensuremath{L_2}}

\newcommand*{\AckInstitutions}{a number of agencies and
  institutes that have supported the development of \Euclid, in
  particular
  the Agenzia Spaziale Italiana,
  the Austrian Forschungsf\"orderungsgesellschaft funded through BMK,
  the Belgian Science Policy,
  the Canadian Euclid Consortium,
  the Deutsches Zentrum f\"ur Luft- und Raumfahrt,
  the DTU Space and the Niels Bohr Institute in Denmark,
  the French Centre National d'Etudes Spatiales,
  the Funda\c{c}\~{a}o para a Ci\^{e}ncia e a Tecnologia,
  the Hungarian Academy of Sciences,
  the Ministerio de Ciencia, Innovaci\'{o}n y Universidades,
  the National Aeronautics and Space Administration,
  the National Astronomical Observatory of Japan,
  the Netherlandse Onderzoekschool Voor Astronomie,
  the Norwegian Space Agency,
  the Research Council of Finland,
  the Romanian Space Agency,
  the State Secretariat for Education, Research, and Innovation (SERI) at the Swiss
  Space Office (SSO),
  and the United Kingdom Space Agency.
  A complete and detailed list is available on the \Euclid\ web site
  (\texttt{http://www.euclid-ec.org}). }

\newcommand{\AckECon}{The authors acknowledge the Euclid Consortium,
  the European Space Agency, and \AckInstitutions}

\title{Tracking radiation damage of \Euclid{} VIS detectors after 1 year in space}

\author[a]{Jesper Skottfelt}
\author[a]{Matt Wander}
\author[b]{Mark Cropper}
\author[a]{Ben Dryer}
\author[a]{David J. Hall}
\author[c]{Richard Hayes}
\author[a]{Bradley Kelman}
\author[b]{Tom Kitching}
\author[d]{Ralf Kohley}
\author[c]{David Lagattuta}
\author[a]{Zoe Lee-Payne}
\author[b]{Patricia Liebing}
\author[c]{Richard Massey}
\author[e]{Henry Joy McCracken}
\author[f]{Reiko Nakajima}
\author[g,c]{James Nightingale}

\affil[a]{Centre for Electronic Imaging, The Open University, Walton Hall, Milton Keynes, MK7~6AA, United~Kingdom}
\affil[b]{Mullard Space Science Laboratory, University College London, Holmbury St Mary, Dorking Surrey RH5~6NT, United~Kingdom}
\affil[c]{Department of Physics, Durham University, South Road, Durham DH1~3LE, United~Kingdom}
\affil[d]{European Space Agency / ESAC, Villanueva de la Cañada, E-28692 Madrid, Spain}
\affil[e]{Institut d'Astrophysique de Paris, 98 bis Boulevard Arago, 75014 Paris, France}
\affil[f]{Argelander-Institut für Astronomie, Auf dem Hügel 71, D-53121 Bonn, Germany}
\affil[g]{School of Mathematics, Statistics and Physics, Newcastle University, Herschel Building, Newcastle-upon-Tyne, NE1 7RU, UK}

\authorinfo{This paper is published on behalf of the Euclid consortium.\\Further author information: (Send correspondence to J.S.)\\J. S.: E-mail: jesper.skottfelt@open.ac.uk, Telephone: +44 (0)1908 652698}

\begin{document} 
\maketitle

\begin{abstract}
Due to the space radiation environment at \Ltwo{}, ESA's \Euclid{} mission will be subject to a large amount of highly energetic particles over its lifetime. These particles can cause damage to the detectors by creating defects in the silicon lattice. These defects degrade the returned image in several ways, one example being a degradation of the Charge Transfer Efficiency, which appears as readout trails in the image data. This can be problematic for the \Euclid{} VIS instrument, which aims to measure the shapes of galaxies to a very high degree of accuracy. Using a special clocking technique called trap pumping, the single defects in the CCDs can be detected and characterised. Being the first instrument in space with this capability, it will provide novel insights into the creation and evolution of radiation-induced defects and give input to the radiation damage correction of the scientific data. We present the status of the radiation damage of the Euclid VIS CCDs and how it has evolved over the first year in space. 
% {\it and evaluate how this compares to the on-ground testing of these devices. [I might not have time to do this]}

\end{abstract}


\section{INTRODUCTION}
\label{sec:intro}  % \label{} allows reference to this section

The main purpose of the VIS instrument\cite{EuclidSkyVIS} on board the European Space Agency’s \Euclid{} mission\cite{EuclidSkyOverview} is to measure the shape of galaxies to very high precision and from that infer the distribution of dark matter in the Universe. 
For this to be possible, all systematic effects need to be well-understood and controlled to a very high degree of accuracy. 

One such systematic effect stems from the radiation damage the spacecraft and imaging sensors will receive at \Ltwo, the 2nd Sun-Earth Lagrangian point 1.5\,million km from Earth. The VIS instrument uses silicon-based charge-coupled devices (CCDs), which can be damaged by highly energetic particles, mainly protons from the Sun or the Galactic plane. When these particles hit the device, they are able to release silicon atoms from the silicon lattice, thus creating defects. 
During integration and readout of the CCDs, these defects can act as traps that can capture photo-electrons and release them at a later point in time, depending on the trap ``species'' and device temperature.
Because of how a CCD is operated, where charge is transferred across the image array through many pixels to be read out, these traps can create trails in the scientific data. 
This effect is known as charge transfer inefficiency (CTI), i.e.~the fraction of charge left behind when it is transferred from one pixel to the next. 
After time in space, this shape change will dominate over the shape distortions due to weak lensing that Euclid is measuring, and it is crucial that the effect of CTI is well understood and can be corrected to very high accuracy. 

For \Euclid{} VIS the Algorithm for CTI Correction (ArCTIC)\cite{Massey2010,Massey2014}  is used to correct CTI in science data. ArCTIC assumes that traps delay electrons stochastically, but with a characteristic time that (in aggregate) leads to trails behind bright objects that are roughly a decaying exponential. The free parameters in this model include the density of traps in a CCD and their distribution of release times. They can be calibrated by fitting the observed trails behind an injected block of charge. However, sums of (nearly) exponential functions are inherently degenerate and the results can be distorted by stray light, bias, other electronic effects\cite{IsraelCTI2015}. There is, therefore, a risk that a non-optimal set of parameters are found. 

To provide additional, independent information about some traps, we also use Trap Pumping (TP; also known as pocket pumping). This was described by Ref.~\citenum{Janesick_2001} as a method to identify pixels with one or more traps in the silicon lattice. This has been further developed by Refs.~\citenum{Hall2014, Hall2017, Skottfelt_interphase, Bush2021} to give more information about the trap species, such as emission time constant and energy level of the trap.

TP was adapted for the VIS CCDs and has been thoroughly tested as part of the on-ground irradiation campaign\cite{Tech_radcamp,C3TM_JATIS_2017,C3TM_SPIE_2018}. 
Here it was shown that getting this extra information about the trap parameters can inform and improve the radiation damage correction. TP was therefore implemented as part of the daily calibration routines that the VIS instrument performs in orbit. 

With the launch of the \Euclid{} telescope on 1 July 2023, the spacecraft has now spent almost a year in space. Here we give a short introduction to the TP method, followed by initial in-orbit trap pumping results and how the trap landscape has evolved over the first year in space. 
%\textit{Finally, we will compare the in-orbit data to the on-ground testing and try and predict how the trap density will increase over the full mission lifetime. }

\section{TRAP PUMPING TECHNIQUE}\label{TP-theory}
TP is a technique that can localise and characterise traps in the silicon lattice down to the electrode, and even sub-electrode level of the single pixels.\cite{Skottfelt_jinst}. 
It works by creating a uniform charge across the entire device, either by illumination with a flat field, or by charge injection. The charge is then shifted a number of clock phases forward and then back to its origin, using the same phase time (\tph) for each step. 
If a trap that has captured an electron, releases it when the charge cloud from the adjacent pixel is closer by, the electron will be ejected and therefore shifted to that pixel. Repeating this shuffle many times will generate a dipole, i.e. a pixel depleted of charge and an adjacent pixel with excess charge compared to the uniform level. From Shockley-Read-Hall theory\cite{Shockley_Read_1952,Hall_1952} it can be found that the intensity (defined as half of the contrast) of this dipole can be determined by the emission time constant (\taue) of the trap and the \tph{} used in the shuffling, using the following equation\cite{Hall2014}:
\begin{equation}
    I = N \cdot \eP \cdot \left[\exp\left(\frac{-\tph}{\taue}\right) - \exp\left(\frac{-2\tph}{\taue}\right) \right] \, , \label{eq:tph}
\end{equation}
where $N$ is the number of pumps and \eP{} is the pumping efficiency, which is related to the capture probability of the trap\cite{Skottfelt_interphase}. 

By repeating this process using different \tph values, a dipole intensity curve as a function of \tph{} can be measured, and by fitting this curve with Equation~\ref{eq:tph}, the best values for \taue{} and \eP{} will be found. 
Doing this process at different temperatures allows a fit to the energy level of the trap species, which can be used to get a deeper understanding of the origin of the trap. However, as the VIS focal plane is kept at a constant temperature, this last step cannot be performed in space but single temperature results can be related to full on-ground testing. 

As an example, the standard clocking scheme for a three-phase device is to clock the charge through phases 1-2-3-1$'$-3-2-1, where 1$'$ denotes the first phase in the next pixel. This will create an ``up-down'' dipole for traps under phase 2 and a ``down-up'' dipole for traps under phase 3, which means that electrode position can be obtained, resulting in sub-pixel trap position. 
To get the traps under phase 1, the clocking scheme can be started at either phase 2 or 3. 

\section{TRAP PUMPING FOR \Euclid{} VIS}
The VIS instrument focal plane comprises of a 6$\times$6 mosaic of CCD273s, which are manufactured by Teledyne-e2v in Chelmsford, UK. Each CCD273 has $\sim$4k\,$\times$\,4k 12\textmu{}m square pixels and four outputs, one in each corner, giving 2048(H)\,$\times$\,2066(V) pixels for each quadrant. They are manufactured on high-resistivity epitaxial silicon and back-thinned to 40\,\um{} thickness\cite{Short_2014}. 

The devices have a charge injection structure in the middle of the device, from which charge can be injected into all 4 quadrants. This is used to inject a uniform signal to be used for the TP technique. 
The amount of charge that is injected depends on the exact geometry of the charge injection electrodes in each column, causing a large column-to-column variation, but the noise on the charge injected within a column is very low. Because of the CCD273 pixel design, more charge is injected in the upper half of the device (quadrants G-H) than the lower half (quadrants E-F).

The pixels have four phases in the image region, or parallel register, designed with alternating widths of 4-2-4-2\,\um{}. The readout, or serial, register pixels have three equal-sized phases with a width of 4\,\um{}.

As TP was originally developed for 3-phase devices, the technique had to be further updated to be used for a 4-phase device, like the parallel direction for the CCD273. 
The resulting clocking scheme is presented in Ref.~\citenum{Skottfelt_jinst} and is named the sub-pixel clocking scheme, as the charge will only see a part of the pixel. 
The scheme works by clocking the charge through phases 1-2-3-2-1, which will only probe the traps in the outer parts of phases 1 and 3, providing not only sub-pixel, but sub-phase positional information about the trap. 
Using four versions of this clocking scheme, each starting in a different phase, the traps in both sides of all four phases can be detected.

TP is part of the in-orbit calibration routines for \textit{Euclid} VIS, which has six daily slots dedicated to TP, however, only one frame can be read out in each slot. During the on-ground calibration campaign, 50 or 100 \tph{} value measurements were used to build the dipole curves for parallel, or vertical, trap pumping (VTP) but running that many \tph{} values in orbit would take 8 to 16 days. 
Taking data for a single dipole curve over such a long period would risk invalidating the data as the trap landscape would change over that time. By running an analysis on many subsets of the on-ground data, it was found that taking 11 \tph{} values between 10\,\us{} and 16\,ms gave the optimal balance between coverage of the dipole curve to avoid fitting errors, keeping within the time allocated to each TP measurement, and taking the data within a reasonable time frame.

It therefore takes less than two days to get a dipole curve for a single set of VTP parameters. With four start phases for the clocking scheme and two different charge injection levels, a full parameter set can be measured in about 15 days.

As the serial register consists of a single row of pixels, serial trap pumping (STP) is done slightly differently. After the charge has been injected into the device, a single line of charge is clocked into the serial register. STP is performed on this line of charge, which is then read out and the next line clocked into the serial register, and so on. 
With the \textit{Euclid} VIS electronics, it is possible to define eight different parameter sets to be read out in a single frame. The serial \tph{} values can therefore be covered with just three TP calibration slots. With two start phases for the serial clocking scheme and two signal levels, only 12 slots are needed to cover all the STP parameters. 
However, due to the need for optimising the serial readout, the VIS electronics is not able to perform STP in the normal way. The output and analysis are therefore slightly different and will not be detailed in this paper. 

%%%%%%%%%%%%%%%%%%%%%%%%%
\section{IN-ORBIT TRAP PUMPING MEASUREMENTS}
\Euclid{} was launched on 1 July 2023 on a Space-X Falcon 9 rocket. It had a 6-week spacecraft commissioning phase, including the two-week transit to \Ltwo, and was then ready to start the Performance Verification (PV) phase. A full set of VTP and STP data was acquired over 12 hours on 15 August as part of PV (in the following referred to as PV1).

\begin{figure}
    \centering
    \includegraphics[width=0.8\linewidth]{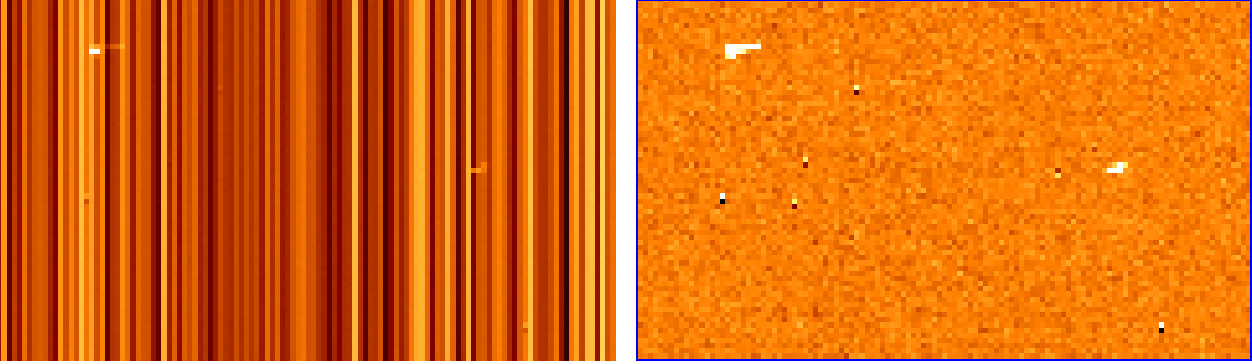}
    \caption{(left) Small section of a raw VTP frame, showing the large column-to-column non-uniformity of the charge injection, and (right) the same section, where the median of each column has been subtracted revealing the dipoles. 
    %The unit of the color bars are electrons, with the zero level in the right-hand side plot set to 20,000\,e$^-$.
    }
    \label{fig:raw_CIsub_RoI}
\end{figure}

\begin{figure}
    \centering
    \includegraphics[width=0.8\linewidth]{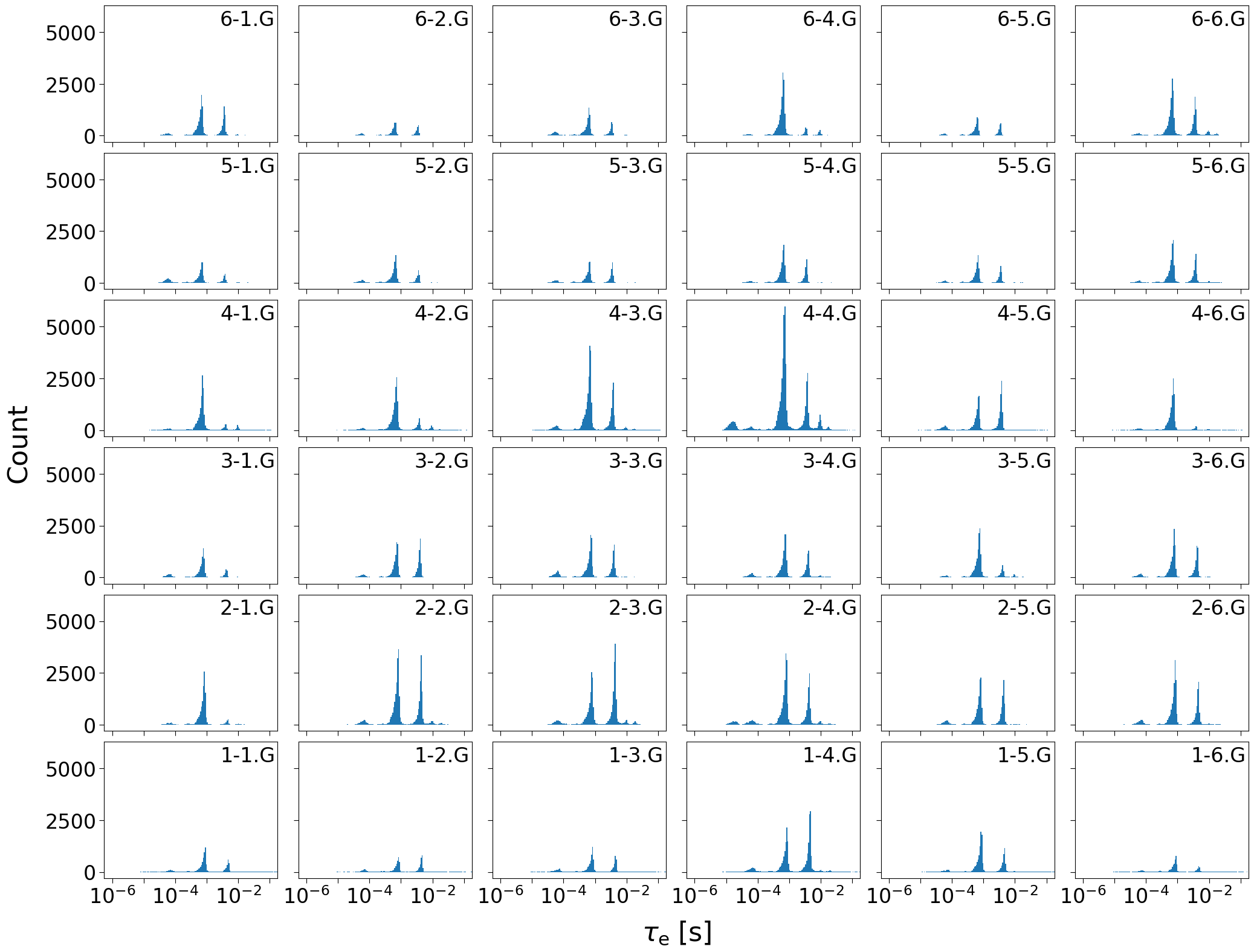}
    \caption{Histograms of \taue values for the G quadrant of all the devices from the PV1 VTP data. The histograms are displayed in the same way as the devices are positioned in the focal plane.}
    \label{fig:taue_hist_PV1}
\end{figure}

A small section of a PV1 VTP raw data frame, and the same section after subtracting the charge injection level for each line, are shown in Fig.~\ref{fig:raw_CIsub_RoI}.
The VTP data was analysed using an implementation of the algorithm described in Sec.~\ref{TP-theory} and a histogram of the detected traps are shown in Fig.~\ref{fig:taue_hist_PV1}. 
This shows all the traps found in one quadrant of the device, by the combined result for all four starting phases for the clocking scheme. 

\begin{figure}
    \centering
    \includegraphics[width=0.8\linewidth]{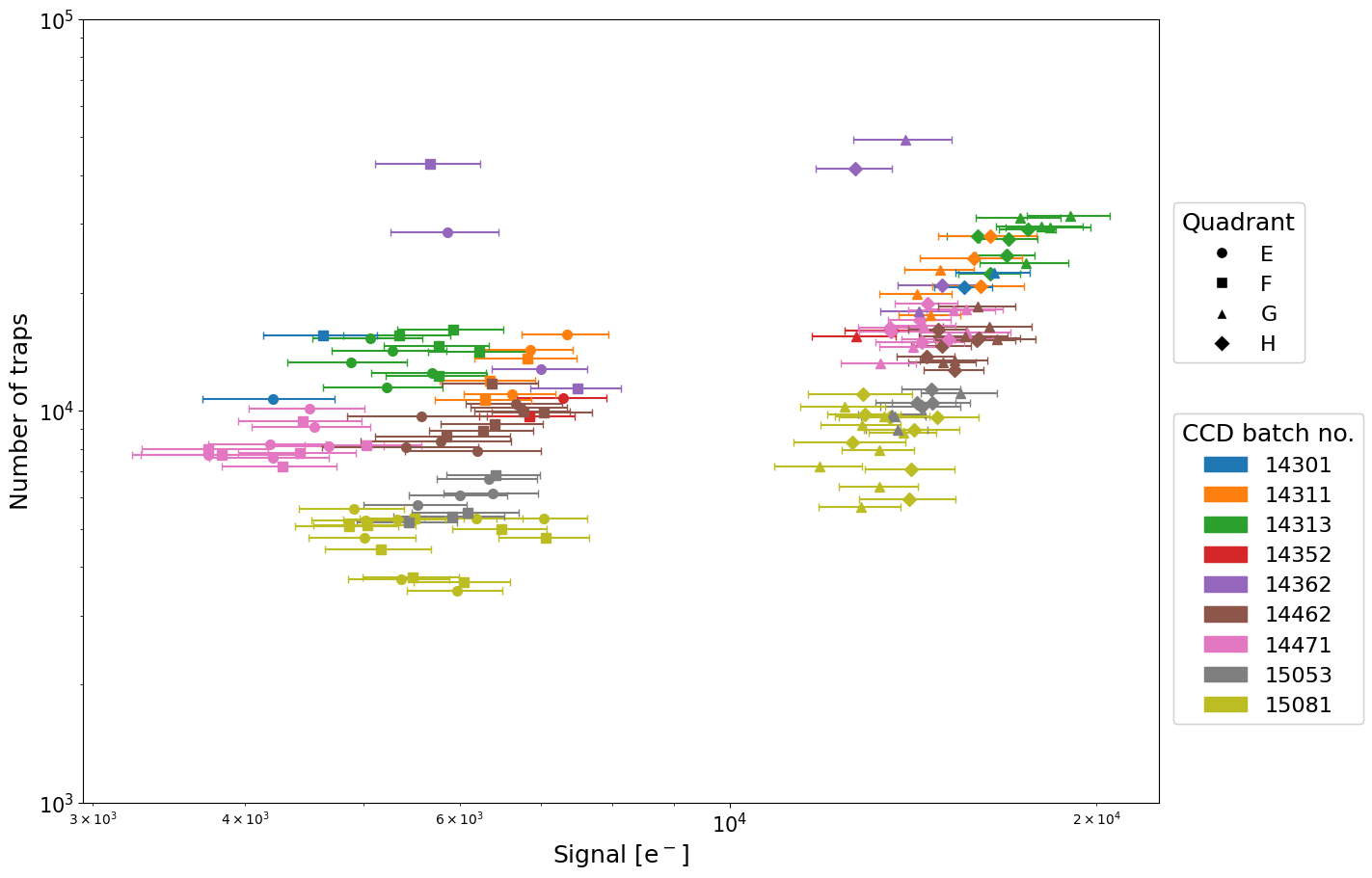}
    \caption{Number of intrinsic traps as a function of signal level, for each of the four quadrants E-H, showing the difference in injection level between the lower half (quadrants E-F) and the upper half (quadrants G-H). The colours represent the silicon wafer batches from which the devices are manufactured. Several CCDs belong to the same batch, which are made from the same bulk silicon. The number of traps detected further depends on the signal level as this changes the size of the charge cloud.}
    \label{fig:ntraps_vs_signal}
\end{figure}

After only 1.5 months in space, it is not expected that the devices have incurred any significant damage and the trap landscape should therefore be very similar to the pre-launch trap landscape.
All of the devices show three distinct peaks at the same location. These peaks are consistent with the pre-irradiation peaks found in on-ground data\cite{Tech_calcampTP}. 
The majority of these are intrinsic, or manufacturing, traps, which are created by the intrinsic Carbon and Oxygen impurities in the silicon wafers, which form stable defects with Phosphorus and Boron atoms used in the doping process of the devices\cite{Skottfelt_jinst}. 
This can be further evidenced by looking at the number of traps in each device and comparing these to the batch numbers of the devices (see Fig.~\ref{fig:ntraps_vs_signal}), which show a clear relationship between the silicon wafers used for the device and the number of intrinsic traps there are. 

On-ground testing has shown that the number of intrinsic traps will be constant across the device and Fig.~\ref{fig:ntraps_vs_signal} therefore also shows that there is a clear correlation between the signal level and the number of traps found in the device. As the E-F quadrants have a lower amount of charge injected than the G-H quadrants, the charge cloud will be smaller in the E-F quadrants and will ``see'' fewer traps than in the G-H quadrants for the same device. 

\begin{figure}
    \centering
    \includegraphics[width=0.8\linewidth]{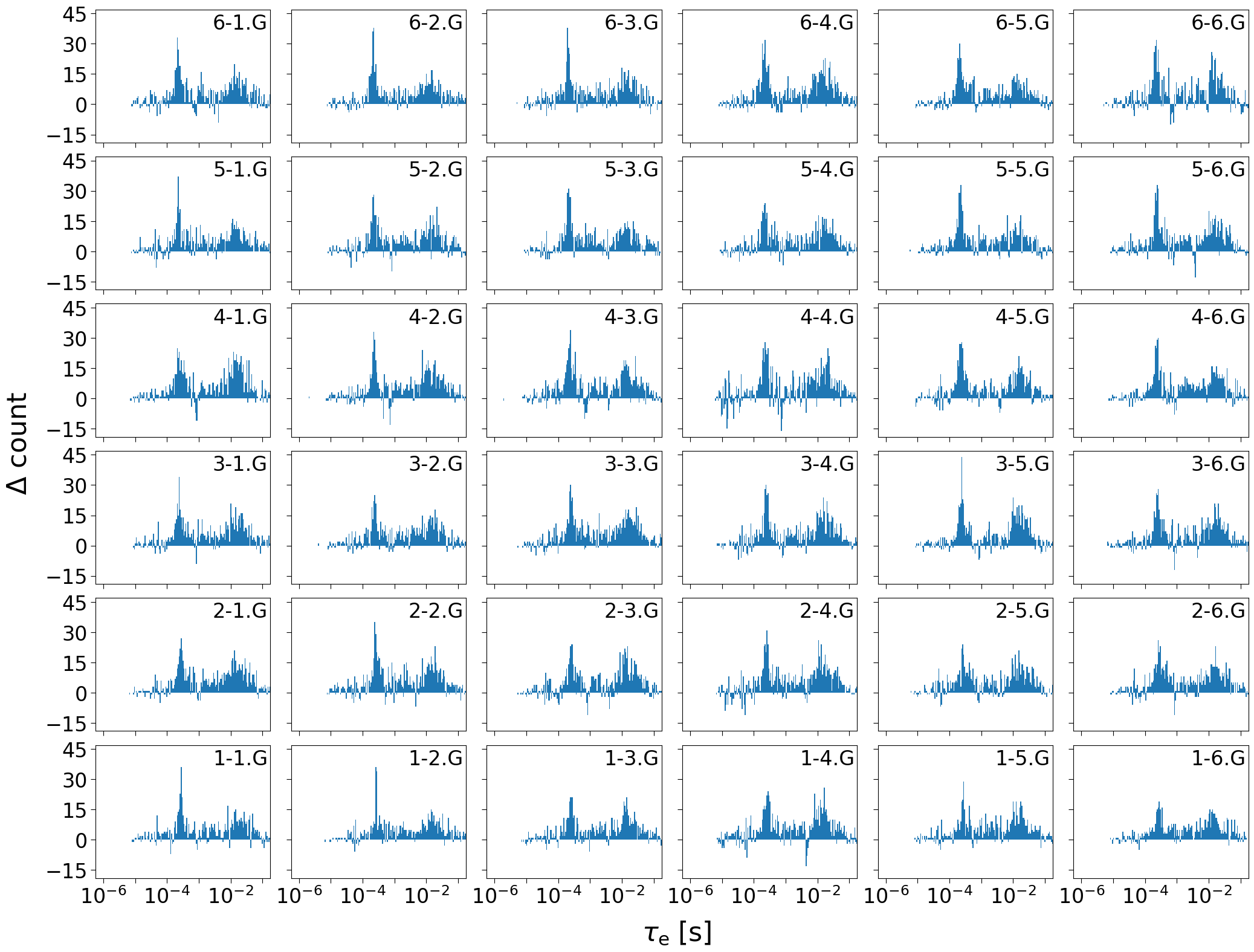}
    \caption{Histograms of difference in traps detected between PV1 (15 August 2023) and PV2 (31 October 2023). }
    \label{fig:PV1_PV2_diff}
\end{figure}

A second full TP parameter data set (PV2) was acquired as part of the PV phase on 31 October, 2.5 months after the first data was obtained. 
%After the data were analysed, the \taue values were corrected for temperature differences between PV1 and PV2. 
Fig.~\ref{fig:PV1_PV2_diff} shows a histogram of the differences between PV1 and PV2. Here traps in the same pixel and phase in the two datasets with $\log(\taue{})$ values within 7\% of each other are taken to be the same trap, to account for inherent noise in the \taue{} measurement (and any temperature difference of the devices between PV1 and PV2). This should remove all the intrinsic traps and only display the traps created in the 2.5 months after the PV1 data was obtained. 

From Fig.~\ref{fig:PV1_PV2_diff} it can be seen that many traps are being created with a continuum of \taue{} values rather than in the well-defined divacancy trap peak\cite{Wood_divacancy} with a \taue{} value of about 220\,\us, at the VIS focal plane operating temperature of about 153\,K. 
This is consistent with the results obtained during the CCD273 irradiation campaign, where the devices were irradiated cold and kept cold for subsequent testing. Here it was found that most traps were created as continuum traps, which over time re-orient in the silicon lattice to a more optimal energy level closer to the divacancy peak. 
Some of the histograms have negative values, which could be attributed to traps created in the short period before PV1 and going through the same process of annealing towards the divacancy peak. 

\begin{figure}
    \centering
    \includegraphics[width=0.8\linewidth]{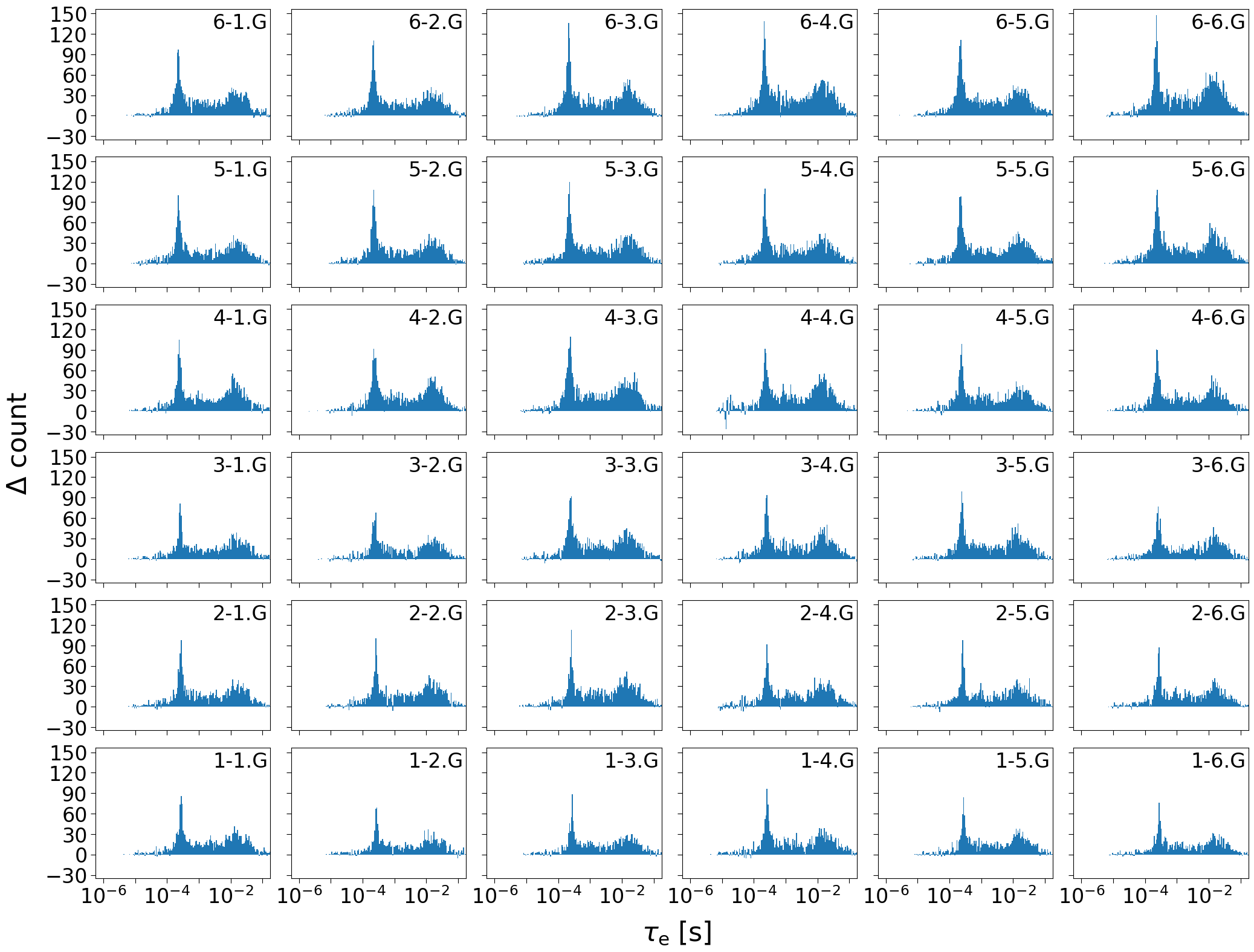}
    \caption{Histograms of difference in traps detected between PV2 (31 October 2023) and end-March 2024. }
    \label{fig:ROS_PV2_diff}
\end{figure}

Since mid-March full TP parameter sets have been obtained through the regular six daily TP measurements and fortnightly updates of the trap densities of the focal plane are therefore available. Fig.~\ref{fig:ROS_PV2_diff} shows the new traps created in the period between PV2 and the end of March 2024. 

This figure looks very similar to the Fig.~\ref{fig:PV1_PV2_diff}, with a distinct divacancy peak at 220\,\us{}, a broad hump around 20\,ms, and a significant number of traps with a continuum of \taue values, which is all consistent with the on-ground radiation damage testing. 
The traps seem to be generated fairly uniformly across the focal plane, maybe with slightly more generated in the upper half, especially the right-hand side, and slightly fewer in the lower part of the focal plane. 
This difference does not seem to be correlated with either the mean injection level of the quadrant or the intrinsic trap density of the device. It could be related to the general shielding of the focal plane or another effect, but the cause for this non-uniformity in trap distribution will become clearer as the total damage increases. 

\begin{figure}
    \centering
    \includegraphics[width=0.8\linewidth]{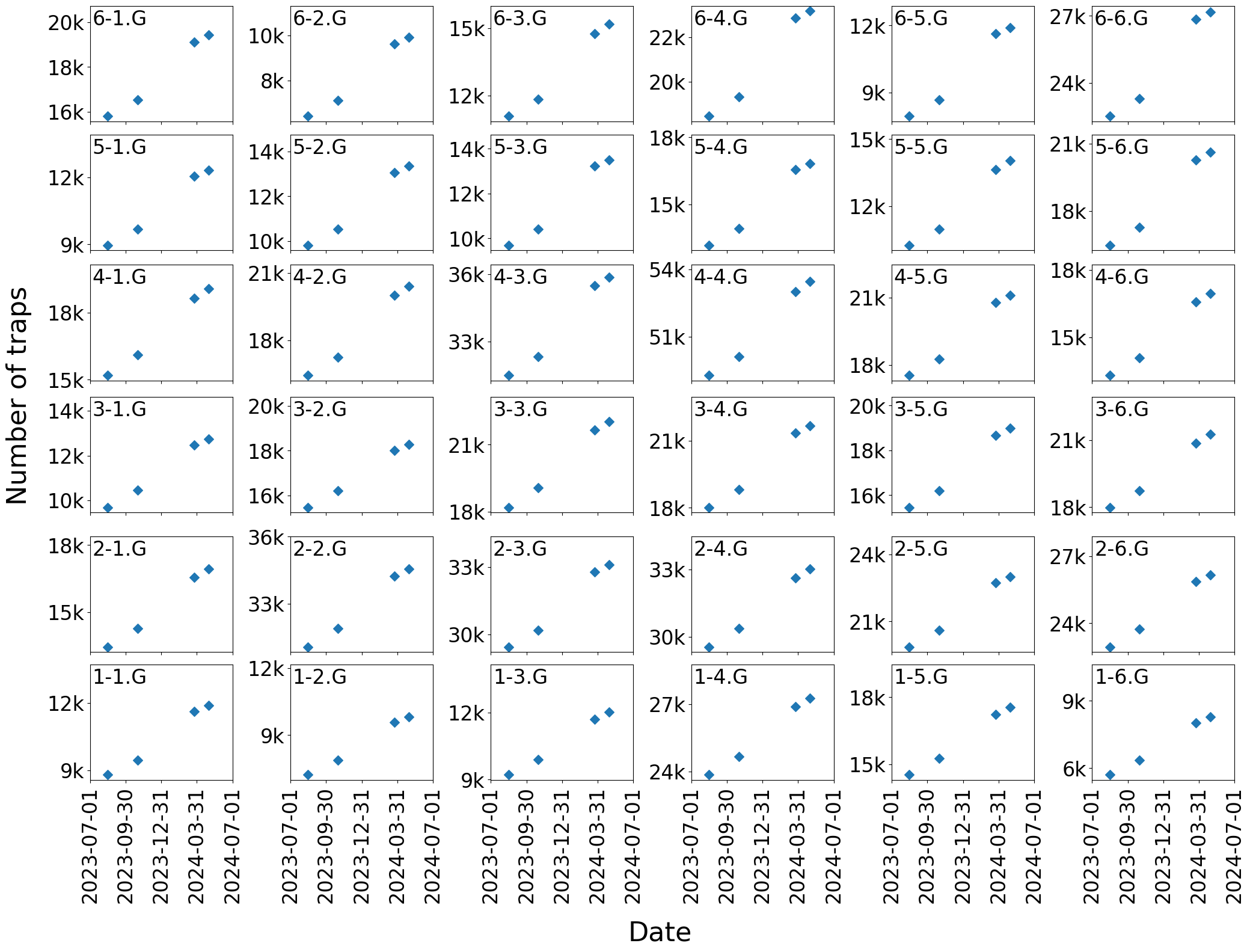}
    \caption{Number of traps found in one quadrant of each device since the launch of \Euclid{}. Note, that the vertical axes of all the sub-plots are scaled equally, making the number of radiation-induced traps in each device directly comparable.}
    \label{fig:traps_v_time}
\end{figure}

The number of new traps generated since the initial PV1 data was obtained, is shown in Fig.~\ref{fig:traps_v_time}. 
While this is early in the mission and only a few data points are available, it shows a clear linear trend of trap generation. However, some of the devices, especially in the upper right-hand side of the focal plane, might be showing indications of jumps. This could be caused by large Coronal Mass Ejections (CMEs), similar to what was found in the Gaia CTI data\cite{Gaia_CTI}, and it will be interesting to track if this behaviour continues.

\section{CONCLUSION}
Radiation damage will have a major impact on the \Euclid{} VIS detectors and it is therefore of great importance that the effect of this damage is corrected for in the VIS data. 
The TP method can detect single defects in the silicon lattice of CCDs, allowing for the detection and tracking of radiation-induced defects in the VIS detectors. 

The initial TP data obtained at the very beginning of the mission shows three distinct species of intrinsic defects. These can be attributed to the intrinsic Carbon and Oxygen impurities in the silicon, which mixed with the Phosphorus and Boron dopants, create stable defects in the manufacturing process.

Subsequent TP data show that traps are created in a continuum and then re-orienting themselves in the silicon lattice to obtain a more optimal energy level closer to the divacancy peak, which matches what was found during the on-ground radiation damage campaign of the CCD273 devices. 
They also show a small variation in trap generation across the focal plane, and that the traps are generated linearly over time, however, with some indication of jumps that could be caused by CMEs. 

The TP method will be performed throughout the \Euclid{} mission and this will provide further measurements of the evolution of damage in the VIS CCDs. The results shown thus far indicate that performing trap pumping in orbit allows the evolution of the trap landscape and therefore CTI of the device to be measured and understood with much more precision than previously possible. This will allow models to correct CTI (e.g. ArCTIC) to be more accurate and therefore meet the science requirements of \Euclid{} for a longer time in the harsh radiation environment in space. Using this technique allows \Euclid{} to perform its science goals at higher irradiation levels, hopefully resulting in a longer useful operating lifetime.

\acknowledgments % equivalent to \section*{ACKNOWLEDGMENTS}     
\AckECon{}

% References
\input{sci_papers}
\bibliography{EuclidTP_1yr} % bibliography data in report.bib
\bibliographystyle{spiebib} % makes bibtex use spiebib.bst

\end{document}

%% file: sci_papers.tex
\def\aj{AJ}%       % Astronomical Journal
\def\actaa{Acta Astron.}%          % Acta Astronomica
\def\araa{ARA\&A}%          % Annual Review of Astron and Astrophys
\def\apj{ApJ}%          % Astrophysical Journal
\def\apjl{ApJ}%          % Astrophysical Journal, Letters
\def\apjs{ApJS}%          % Astrophysical Journal, Supplement
\def\ao{Appl.~Opt.}%          % Applied Optics
\def\apss{Ap\&SS}%          % Astrophysics and Space Science
\def\aap{A\&A}%          % Astronomy and Astrophysics
\def\aapr{A\&A~Rev.}%          % Astronomy and Astrophysics Reviews
\def\aaps{A\&AS}%          % Astronomy and Astrophysics, Supplement
\def\azh{AZh}%          % Astronomicheskii Zhurnal
\def\baas{BAAS}%          % Bulletin of the AAS
\def\bac{Bull. astr. Inst. Czechosl.}%          % Bulletin of the Astronomical Institutes of Czechoslovakia 
\def\caa{Chinese Astron. Astrophys.}%          % Chinese Astronomy and Astrophysics
\def\cjaa{Chinese J. Astron. Astrophys.}%          % Chinese Journal of Astronomy and Astrophysics
\def\icarus{Icarus}%          % Icarus
\def\jcap{J. Cosmology Astropart. Phys.}%          % Journal of Cosmology and Astroparticle Physics
\def\jrasc{JRASC}%          % Journal of the RAS of Canada
\def\mnras{MNRAS}%          % Monthly Notices of the RAS
\def\memras{MmRAS}%          % Memoirs of the RAS
\def\na{New A}%          % New Astronomy
\def\nar{New A Rev.}%          % New Astronomy Review
\def\pasa{PASA}%          % Publications of the Astron. Soc. of Australia
\def\pra{Phys.~Rev.~A}%          % Physical Review A: General Physics
\def\prb{Phys.~Rev.~B}%          % Physical Review B: Solid State
\def\prc{Phys.~Rev.~C}%          % Physical Review C
\def\prd{Phys.~Rev.~D}%          % Physical Review D
\def\pre{Phys.~Rev.~E}%          % Physical Review E
\def\prl{Phys.~Rev.~Lett.}%          % Physical Review Letters
\def\pasp{PASP}%          % Publications of the ASP
\def\pasj{PASJ}%          % Publications of the ASJ
\def\qjras{QJRAS}%          % Quarterly Journal of the RAS
\def\rmxaa{Rev. Mexicana Astron. Astrofis.}%          % Revista Mexicana de Astronomia y Astrofisica
\def\skytel{S\&T}%          % Sky and Telescope
\def\solphys{Sol.~Phys.}%          % Solar Physics
\def\sovast{Soviet~Ast.}%          % Soviet Astronomy
\def\ssr{Space~Sci.~Rev.}%          % Space Science Reviews
\def\zap{ZAp}%          % Zeitschrift fuer Astrophysik
\def\nat{Nature}%          % Nature
\def\iaucirc{IAU~Circ.}%          % IAU Cirulars
\def\aplett{Astrophys.~Lett.}%          % Astrophysics Letters
\def\apspr{Astrophys.~Space~Phys.~Res.}%          % Astrophysics Space Physics Research
\def\bain{Bull.~Astron.~Inst.~Netherlands}%          % Bulletin Astronomical Institute of the Netherlands
\def\fcp{Fund.~Cosmic~Phys.}%          % Fundamental Cosmic Physics
\def\gca{Geochim.~Cosmochim.~Acta}%          % Geochimica Cosmochimica Acta
\def\grl{Geophys.~Res.~Lett.}%          % Geophysics Research Letters
\def\jcp{J.~Chem.~Phys.}%          % Journal of Chemical Physics
\def\jgr{J.~Geophys.~Res.}%          % Journal of Geophysics Research
\def\jqsrt{J.~Quant.~Spec.~Radiat.~Transf.}%          % Journal of Quantitiative Spectroscopy and Radiative Trasfer
\def\memsai{Mem.~Soc.~Astron.~Italiana}%          % Mem. Societa Astronomica Italiana
\def\nphysa{Nucl.~Phys.~A}%          % Nuclear Physics A
\def\physrep{Phys.~Rep.}%          % Physics Reports
\def\physscr{Phys.~Scr}%          % Physica Scripta
\def\planss{Planet.~Space~Sci.}%          % Planetary Space Science
\def\procspie{Proc.~SPIE}%% Proceedings of the SPIE
\let\astap=\aap
\let\apjlett=\apjl
\let\apjsupp=\apjs
\let\applopt=\ao

\def\mdash{--}